\documentstyle[12pt]{article}
\def\NPB{{\it Nucl. Phys. }{\bf B}}
\def\PL{{\it Phys. Lett. }}
\def\PRL{{\it Phys. Rev. Lett. }}

\def\IJMPA{{\it Int. J. Mod. Phys. }{\bf A}}

\def\PR{{\it Phys. Rep. }}

\newcommand{\eq}{\begin{equation}}
\newcommand{\en}{\end{equation}}
\newcommand{\eqn}{\begin{eqnarray}}
\newcommand{\enn}{\end{eqnarray}}
\newcommand{\nn}{\nonumber }
\newcommand{\beq}{\begin{equation}}
\newcommand{\eeq}{\end{equation}}

\let\ba=\overline
\def\CP#1{\relax\ifmmode\IP^{#1}\else\IP$^{#1}$\fi}
\def\define{\buildrel{\rm def}\over=}
\let\f=\phi
\let\a=\alpha
\def\Imm{\Im m}
\def\inv#1{{\textstyle{1\over#1}}}
\def\IP{\relax\leavevmode{\rm I\kern-.18em P}}
\let\q=\theta
\let\p=\pi
\def\Ree{\Re e}
\let\t=\tau
\let\To=\Rightarrow
\let\vd=\partial
\let\W=\Omega
\def\zb{\bar{z}}
\def\chs#1#2{{\textstyle{#1\choose#2}}}
\def\frc#1#2{{\textstyle{#1\over#2}}}
\def\IR{\relax\leavevmode{\rm I\kern-.18em R}}
\def\ZZ{\relax\leavevmode
           \ifmmode\mathchoice
           {\hbox{\sf Z\kern-.4em Z}}
           {\hbox{\sf Z\kern-.4em Z}}
           {\lower.9pt\hbox{\scriptsize\sf Z\kern-.36em Z}}
           {\lower1.2pt\hbox{\tiny\sf Z\kern-.36em Z}}
            \else{\sf Z\kern-.4em Z}\fi}

\thicklines     

\begin{document}
\input epsf.tex

\begin{titlepage}
\begin{flushright}
NSF-ITP-00-37 \\
CITUSC/00-022 \\
hep-th/0005162\\
\end{flushright}

\begin{center}

{\Large\bf EXPONENTIAL HIERARCHY FROM\\[2mm]
        SPACETIME VARIABLE STRING VACUA} \\[10mm]
{\bf P. Berglund\footnote{e-mail: berglund@itp.ucsb.edu} } \\[1mm]
    Institute for Theoretical Physics\\
    University of California, Santa Barbara\\
    Santa Barbara, CA 93106\\[5mm]
{\bf T. H\"{u}bsch\footnote{e-mail: thubsch@howard.edu}%
    $^,$\footnote{On leave from the ``Rudjer Bo\v skovi\'c'' Institute,
              Zagreb, Croatia.} } \\[1mm]
    Department of Physics and Astronomy\\
    Howard University\\
    Washington, DC 20059\\[5mm]
{\bf D. Minic\footnote{e-mail: minic@physics.usc.edu} } \\[1mm]
    CIT-USC Center for Theoretical Physics\\
    Department of Physics and Astronomy\\
    University of Southern California\\
    Los Angeles, CA 90089-0484\\[10mm]
{\bf ABSTRACT}\\[3mm]
\parbox{4.5in}{It is shown that
non-supersymmetric spacetime varying string vacua 
can lead to an exponential hierarchy
between the electroweak and the gravitational scales. The hierarchy is
naturally generated by a string coupling of $O(1)$.}
\end{center}

\end{titlepage}

\section{Introduction}
The possibility of our $3{+}1$-dimensional world being a cosmic
defect (brane) in a higher-dimensional theory~\cite{rubakov, visser,
anton, georgia, horava, braneTH, savas, branew1, branew2, branew3,
rs1, rs2}
has recently attracted much interest. In particular, the observed
hierarchy between the electroweak and the gravitational scales was
considered in this context in~\cite{rs1}.

In this paper, we show how a class of non-supersymmetric
string vacua can naturally lead to an exponential
hierarchy. In general we consider $p{+}1$-dimensional cosmic defects
embedded in a $D$-dimensional spacetime. These types of models have been 
considered before in the literature~\cite{cohen, cp, RG, tony}. Here
we show explicitly that such cosmic defects emerge as spacetime
varying string vacua. The exponential hierarchy  between
the electroweak and gravitational scales arises from
non-trivial warp factors in the metric 
and naturally is generated by the string coupling of $O(1)$.
(The r\^ole of warp factors in string theory and their relationship to
cosmic brane models
have been studied  in~\cite{cvetic, verlinde, greene, mayr}.)

Our solutions resemble the stringy cosmic
strings~\footnote{Historically these solutions were studied in four
dimensions in which the defects correspond to string like objects. 
In the remainder of this article 
we will use the notation cosmic brane as
this is more appropriate for the current context.}
of~\cite{vafa, gh}. The general framework, described in section~2, is
that of a higher-dimensional string theory compactified on a
Calabi-Yau (complex) $n$-fold, $M_n$, some moduli of which are
allowed to vary over part of the noncompact space. In the
uncompactified Type~IIB theory, the r\^ole of space-dependent moduli
is played by the dilaton-axion system, very much like in Vafa's
description of F-theory~\cite{FTh}. The cosmic defect (brane)
appears as a singularity of the induced spacetime metric, with its
characteristics governed by the energy momentum
tensor of the moduli. 
(In addition, a naked singularity is located 
at a finite proper distance from the core of the brane.)
However, our solutions are non-BPS. This points out to
possible stability problems whose
detailed study we defer to a future work~\cite{bhmtwo}.

The paper is organized as follows:
    In section~2, we recall the construction of spacetime
varying string vacua and generalize
the Ans\"atze of Refs.~\cite{vafa, gh, FTh} for the
metric; we find non-trivial solutions in which the moduli are
non-holomorphic functions over the noncompact space. While this
breaks supersymmetry, the exponential warp factors in the metric
induce a large hierarchy between the Planck scales in the higher- and
lower-dimensional spacetime.
    In section~3, we study our spacetime-varying dilaton-axion solution
when further compactified on a $K3$ or $T^4$ to four dimensions,
and the generation of an
exponential hierarchy.  We briefly consider the case of
compactifying string theory on a spacetime-varying $K3$, which also
gives rise to $3+1$-dimensional vacua with exponential hierarchy.
    Finally, in section~4, we discuss various
possible generalizations of our present work.

\section{Spacetime Varying Vacua and Warp Factors}
Let us consider compactifications of string theory in which the
``internal'' space (a Calabi-Yau $n$-fold $M_n$) 
varies over the ``observable'' spacetime. The parameters of the
``internal'' space then become spacetime variable moduli fields
$\f^\a$. The effective action describing the coupling of
moduli to gravity of the observable spacetime
can be derived by dimensionally reducing the higher dimensional
Einstein-Hilbert action~\cite{vafa,gh}. In this procedure one retains
the dependence of the Ricci scalar in the gravitational action only
on the moduli $\f^\a$. Then, the relevant part of the low-energy
effective $D$-dimensional action of the moduli of the Calabi-Yau
$n$-fold, $M_n$, coupled to gravity reads~\footnote{As a concrete
example, the reader may think of a Type~II string theory compactified
on a $T^2$, in which case $D{=}8$ and $n{=}1$, or on a $K3$ or $T^4$,
when $D{=}6$ and $n{=}2$, {\it etc}.}
\begin{equation}
    S_{\rm eff} = \int d^D x \sqrt{-g} ( -\inv2 R
            - \inv2
G_{\a \bar{\beta}}g^{\mu \nu}
              \vd_{\mu} \f^\a \vd_{\nu} \f^{\bar{\beta}}
              +...)~,
\end{equation}
where $\mu,\nu=0,{\cdots},D-1$. We neglect higher derivative
terms and set the other fields in the theory to zero as in~\cite{vafa}.

We will restrict the moduli to depend 
on $x_i$,
$i{=}D{-}2,D{-}1$, so that $\vd_a \f{=}0$,
$a{=}0,{\cdots},D{-}3$.  The equations of motion are
\begin{equation}
    g^{ij} \Big(\nabla_{i} \nabla_{j} \f^\a
    + \Gamma^\a_{\beta \gamma} (\f, \bar{\f}) \vd _{i}
    \f^{\beta} \vd_{j} \f^{\gamma} \Big) = 0~,\label{e:scalar}
\end{equation}
and
\begin{equation}
    R_{\mu\nu} - \inv2 g_{\mu\nu} R = T_{\mu\nu} (\f,\bar{\f})~,
\label{e:einstein}
\end{equation}
where the energy-momentum tensor of the moduli is
\begin{equation}
    T_{\mu \nu}
    = - G_{\a \bar{\beta}} \Big(\vd_{\mu} \f^\a\vd_{\nu}
     \f^{\bar{\beta}} - \inv2 g_{\mu \nu}\, g^{\rho\sigma}
     \vd_{\rho} \f^\a \vd_{\sigma} \f^{\bar{\beta}}\Big)~.
\end{equation}

It is useful to define $z \equiv (x_{D-2} + i x_{D-1})$ and
rewrite the effective action  as
\begin{equation}
    S_{\rm eff} = -\inv2\int d^D x~ \sqrt{-g}\> R
                  - \inv2\int d^{D-2}x~ E
\end{equation}
where $d^{D-2}x$ refers to the integration measure over the first
$D-2$ coordinates, $x_0,{\cdots},x_{D-3}$. $E$, which we later interpret 
as the energy density
(tension) of the cosmic brane, is given by
\begin{equation}
    E \equiv \int d^2z \sqrt{-g}~ G_{\a \bar{\beta}} g^{z \bar{z}}
    \Big(\vd_{z} \f^\a \vd_{\bar{z}} \f^{\bar{\beta}}
     + \vd_{\bar{z}}\f^\a\vd_{{z}}
     \f^{\bar{\beta}}\Big)~.\label{e:Etension}
\end{equation}

In order to solve the coupled equations of motion for the moduli and
gravity, Eqs.~(\ref{e:scalar}) and~(\ref{e:einstein}), we start with
the following Ansatz for the metric
\begin{equation}
    ds^2 = e^{2 A(z, \bar{z})} \eta_{ab} dx^a dx^b
    + e^{2B(z, \bar{z})} dz d\bar{z}~.\label{e:metric}
\end{equation}
This type of Ansatz has appeared recently in various field
theory~\cite{cohen,cp}, supergravity inspired
scenarios~\cite{CHT} and in the context of string theory~\cite{ein,dudas}.
The authors in~\cite{CHT} considered the possibility of having a
superpotential and hence a potential for the scalar fields. The
scalars in our effective action are assumed to be true moduli, with
no (super)potential.

Using this Ansatz, Eqs.~(\ref{e:scalar}) and~(\ref{e:einstein})
produce the following~\cite{CHT}.
    The `$zz$' Einstein equation leads to ($\vd \equiv \vd_{z}$ and
$\bar{\vd} \equiv \vd_{\bar{z}}$)
\begin{equation}
    (D{-}2)\Big[ 2 \vd A {\vd} B -  (\vd A)^2 -  {\vd}^2 A\Big] =
    2G_{\a \bar{\beta}}\vd \f^\a {\vd} \f^{\bar{\beta}}
    \label{e:zz}
\end{equation}
and the `$z\bar{z}$' Einstein equation becomes
\begin{equation}
   (D{-}2)\Big[\vd \bar{\vd} A
   + (D{-}2) \vd A \bar{\vd} A\Big] =0.
   \label{e:zbarz}
\end{equation}
Note that the `$\bar z\bar z$' equation is obtained straightforwardly
from~(\ref{e:zz}) by replacing $\vd$ by $\bar\vd$.
We consider the case when $D{>}2$. Then, Eq.~(\ref{e:zbarz}) lets us
simplify the `$ab$' Einstein equation into
\begin{equation}
    (D{-}3)\vd \bar{\vd} A + 2\vd \bar{\vd} B
    = - G_{\a \bar{\beta}}(\vd
    \f^\a \bar{\vd} \f^{\bar{\beta}} +\bar{\vd} \f^\a
    {\vd} \f^{\bar{\beta}})~.\label{e:ab}
\end{equation}
Finally, the equation for the moduli reads
\begin{equation}
   (D{-}2)\Big[2 \vd A \bar{\vd} \f^\a
   + 2 \vd \f^\a \bar{\vd} A\Big] + 4\vd \bar{\vd}
   \f^\a + 4G^{\a \bar{\gamma}} \frac{\vd G_{\delta
   \bar\gamma}}{\vd \f^{{\beta}}}
   \vd \f^{\beta} \bar{\vd} \f^{\delta}=0~.\label{e:moduli}
\end{equation}

Let us start with the well-known supersymmetric solution 
by requiring that $A{=}0$
and $\bar\vd\f^\a{=}0$~\cite{vafa,gh}. Because of holomorphicity, we can
simplify the energy density~(\ref{e:Etension}). In the case of a
variable Calabi-Yau $n$-fold $M_n$ 
\begin{equation}
   G_{\a \bar{\beta}} \equiv \vd_\a \vd_{\beta} K,
   \quad K \equiv -\log{( i^n \int \Omega \wedge\bar{\Omega})}~,
\end{equation}
where $\Omega$ is the appropriate holomorphic $n$-form on $M_n$.
Therefore, Eq.~(\ref{e:Etension}) simplifies to
\begin{equation}
   E = - i \int\vd\bar{\vd}
    \log{( i^n\int\Omega\wedge\bar{\Omega})}~.\label{e:Eholomorphn}
\end{equation}
Note that $\sqrt{-g}g^{z\bar z}=1$ because $A=0$. Finally, one can
show that the tension is equal to the deficit angle, $\delta\theta$, 
caused by the cosmic brane~\cite{vafa,gh}. As expected this saturates
the BPS bound,
\begin{equation}
  E\geq \delta\theta~.\label{e:BPS}
\end{equation}

In order to explicitly solve the equations of motion we consider
$M_1=T^2$, fix the K\"ahler structure 
and focus on the complex structure modulus,
$\tau$, for which the metric on the
moduli space is
\begin{equation}
G_{\tau\bar\tau}=\frac{-1}{(\tau-\bar\tau)^2}~.\label{e:taumetric}
\end{equation}
This is the case considered in~\cite{vafa}.
There, the
authors studied a Type~II string theory compactified on either
a {\it fixed}
$K3$ or $T^4$, followed by a further $T^2_z$ compactification from
six to four dimensions. The subscript $z$ indicates that the torus
varies over two ($x_{2,3}$) of the remaining four coordinates,
$x_i,\, i=0,...,3$.
    When the moduli of $T^2_z$ depend holomorphically on $z=x_2{+}ix_3$
cosmic brane solutions with finite energy density which saturate the
BPS-bound~(\ref{e:BPS}), emerge at particular points in the
$z$-plane. Since $\tau=\tau(z)$ Eq.~(\ref{e:moduli}) 
for the  modulus $\tau$ is immediately satisfied. 
The only remaining equation, (\ref{e:ab}),  reads
\begin{equation}
   \vd \bar{\vd} (2B) = \vd \bar{\vd}
   \log{(\tau{-}\bar\tau)}
\end{equation}
which can be explicitly solved and gives rise to the stringy cosmic
branes~\cite{vafa}.

We are however interested in other solutions for which $A,B{\neq}0$,
but still $\bar\vd\f^\a{=}0$. Although the $\f^\a$ are
holomorphic, this does not necessarily guarantee that the solution is
supersymmetric~\cite{ein}.
(For example, the general analysis of the supersymmetry
transformations in $D=10$ type IIB supergravity implies that
supersymmetry is broken if the modulus $\tau$, the axion-dilaton
system, is non-holomorphic, or the warp factor
$A{\neq}0$~\cite{ein}.)
    The holomorphicity of $\f^\a$ simplifies the equations of
motion considerably. In particular, from Eq.~(\ref{e:moduli}) we
have that moreover $\f^\a=${\it const}, and one finds the
general solution:
\begin{eqnarray}
     e^{2A}&=&\Big({a(z)+\bar{a}(\bar z)\over2a_0}\Big)^{2\over D-2}~,
    \label{e:expA}\\
     e^{2B}&=&|b_0|^2
      \Big({2a_0\over a(z)+\bar{a}(\bar z))}\Big)^{D-3\over D-2}
     |\vd a(z)|^2~, \label{e:expB}
\end{eqnarray}
where $a_0$ and $b_0$ are integration constants and $a(z)$ is an
{\it arbitrary\/} holomorphic function.
Because $\f^\a=${\it const}
the geometry of
the moduli has decoupled.
    We therefore turn to non-supersymmetric solutions with
non-holomorphic moduli. In order to simplify the problem we
look for an axially symmetric solution in the plane perpendicular to
the cosmic brane.
The metric~(\ref{e:metric}) then becomes
\begin{equation}
    ds^2 = e^{2 A(\rho)} \eta_{ab}\, dx^a dx^b
    + \ell^2 e^{2\tilde B(\rho)} \Big(d\rho^2 + d\q^2\Big)~,
    \label{e:metricrho}
\end{equation}
where $z=r_0 e^{\rho_0-\rho+i\q}$ and $\tilde B(\rho)=B(z,\bar z) - \rho$;
the constant parameter $\ell=r_0 \exp(\rho_0)$
has dimensions of length.
Note that this change of variables corresponds to choosing
$a(z)=-\log(z/\ell)$ and $b_0=\ell\tilde b_0$
in Eqs.~(\ref{e:expA}) and~(\ref{e:expB}).
The Einstein equations~(\ref{e:zz})-(\ref{e:ab}) then become
\begin{eqnarray}
    2\Big[\chs{D-2}{2}(A')^2 + (D{-}3)A'' + \tilde B''\Big]
    &=& - G_{\a\bar\beta}(\vd_\rho\f^\a
    \vd_\rho \f^{\bar\beta} + \vd_\q
    \f^\a \vd_\q \f^{\bar\beta})~,\label{e:abrho}\\
    2(D{-}2)\Big[\frc{(D{-}3)}{2}(A')^2 +A'\tilde B'\Big]
    &=& +G_{\a\bar\beta}(\vd_\rho \f^\a \vd_\rho\f^{\bar\beta}
    -\vd_\q \f^\a \vd_\q \f^{\bar\beta})~,
    \label{e:rhorho}\\
    2(D{-}2)\Big[\frc{(D{-}1)}{2}(A')^2 + A'' - A' \tilde B' \Big]
    &=& -G_{\a\bar\beta}(\vd_\rho \f^\a {\vd}_\rho\f^{\bar\beta}
    -\vd_\q \f^\a {\vd}_\q \f^{\bar\beta})~,
    \label{e:thth}
\end{eqnarray}
while the equation for the moduli~(\ref{e:moduli}) reads
\begin{equation}
    2(D{-}2)A'\vd_\rho\f^\a
    + 2(\vd^2_\rho+\vd^2_\q)\f^\a +
   2G^{\a \bar{\gamma}} \frac{\vd G_{\delta
   \bar\gamma}}{\vd \f^{{\beta}}}
   (\vd_\rho \f^{\beta} {\vd_\rho} \f^{\delta}+
   \vd_\q \f^{\beta} {\vd_\q}
   \f^{\delta})=0~.\label{e:modulirho}
\end{equation}
We have used the fact that the two warp factors, $A,\tilde B$, in
Eq.~(\ref{e:metricrho}) depend only on $\rho$, and abbreviated
$A'\equiv\vd_\rho{A}$, {\it etc}. 

To solve these equations we will consider a single modulus scenario,
$\f^\a{=}\tau$ in which $\tau=a + ie^{-\f}$, {\it i.e.}, the
axion-dilaton system of the $D{=}10$ Type~IIB string theory. The
holomorphic solution $\tau=\tau(z)$ is that of
$D7$-branes~\cite{FTh}. A more general solution can be obtained
along the lines of our earlier analysis. In terms of the $\rho$ and
$\q$ variables we have for non-holomorphic
$\tau$ and $A,\tilde B\neq 0$ the following solution,
\begin{eqnarray}
   e^{2A(\rho)} &=& ({\rho\over\rho_0})^{{2\over D-2}}
   \label{e:A}\\
   e^{2\tilde B(\rho)} &=&
   ({\rho_0\over\rho})^{D-3\over D-2}e^{(\rho_0^2-\rho^2)/\rho_0}~.
   \label{e:Bone}
\end{eqnarray}
This is obtained by requiring that $\vd_\rho\tau=0$
in~(\ref{e:modulirho})~\footnote{If $\vd_\theta\tau=0$ one
obtains a different solution without the factor
$\exp((\rho_0^2-\rho^2)/\rho_0)$ essentially reproducing~(\ref{e:expB})
but with a different power of $\rho$~\cite{ein}. The modulus turns out to be
functionally the same as $\tau_{II}$ in (28) below.}.

We find two solutions for $\tau$, depending on whether
$\tau$ is purely imaginary or not~\footnote{The functional form of our
solutions for the
dilaton-axion system
resembles  somewhat the non-supersymmetric electric $D7$-branes
solution of IIB supergravity \cite{Gibbons}. The actual form of the
metric describing our solution in ten dimensions is  also similar to
the recently  found classical background of the non-supersymmetric
$USP(32)$  Type I string theory \cite{dudas}.}. Also, note that
Eqs.~(\ref{e:abrho})--(\ref{e:modulirho}) are unchanged if
$\f^\a{=}\tau$ is rescaled by a positive multiplicative (normalization)
constant.
   Restricting to $\tau=-\bar\tau$, we find the simple particular
solution
\begin{equation}
   \tau_I(\q) = a_0 + i g_0^{-1}e^{2\sqrt{2}\q\over\sqrt{\rho_0}}~,\qquad
   \theta \in [-\pi, \pi]~.
   \label{e:tau1}
\end{equation}
Note that when $a_0\neq 0$ $\tau_I$ satisfies~(\ref{e:modulirho}) because of the form the metric $G_{\tau\bar\tau}$ takes~(\ref{e:taumetric}).
The constants $a_0$ and $g_0$ are chosen such that
rotations through the defining domain of $\q$ induce an $SL(2,\ZZ)$
action on this solution:
\begin{equation}
     \tau_I(+\pi)=\frac{a\tau(-\pi))+b}{-\tau_I(-\pi)+d}~. \label{e:tau1tr}
\end{equation}
With $\tau\neq-\bar\tau$, we find the following particular solution
\begin{equation}
   \tau_{II}(\q) =
   \Big(\frac{n}{2}{\textstyle\coth[{2\sqrt{2}\pi\over\sqrt{\rho_0}}]}\Big)
   \frac{\pm\sinh[{2\sqrt{2}\q\over\sqrt{\rho_0}}]+i}
   {\cosh[{2\sqrt{2}\q\over\sqrt{\rho_0}}]}~,\qquad
   \theta \in [-\pi, \pi]~,
\label{e:tau2}
\end{equation}
where both the axion and the dilaton have a more complicated
$\q$-dependence. Rotations through the defining domain of $\q$
induce a monodromy action on this solution:
\begin{equation}
     \tau_{II}(+\pi)=\tau_{II}(-\pi)\pm n~. \label{e:tau2tr}
\end{equation}


Finally, before turning to a more detailed analysis of these solutions,
we compute the RR-charge. For~(\ref{e:tau1}) the charge is obviously zero as
the axion is a constant. This is an indication that this
particular cosmic brane is not a $D$-brane~\footnote{It may well
turn out, however, that we instead have
a $D$--$\bar D$-brane system. We thank N.~Itzhaki for pointing
this out to us.}. Still, there are certain $D7$-brane configurations
whose RR-charge are zero~\cite{zwiebach}.
For~(\ref{e:tau2}) we note that when $\rho_0>>8\pi^2$ we can write
the axion as $a\sim \pm n\theta/(2\pi)$ from which we
immediately read off the RR-charge as $Q=\pm n$. (In
fact, it is easy to show that this holds for all values of $\rho_0$.)
In this limit we also note that to leading order $\exp(-\phi)=1$.
For a fixed $\ell$, {\it i.e.}, a fixed stress-energy tensor,
$\rho_0\to\infty$ corresponds to taking
$r_0\to 0$ and hence the
core to zero size (see below for identification of the core in our solution).
We thus obtain a situation which is more familiar from that of a
$D7$-brane, except for the existence of the singularities at $\rho=0$
and $\rho=\infty$. One may hope that string corrections may render
these singularities harmless in such a way that the relation between our
solutions and the standard $D7$-branes can be made clearer (see also
the discussion in section~3). We will
defer a more detailed study of these matters for future work~\cite{bhmtwo}.

\section{Exponential Hierarchy}

In the model considered above, where the dilaton and axion vary over
$z{=}x_8{+}ix_9$, 
the metric
\begin{equation}
    ds^2 = (\frac{\rho}{\rho_0})^{{2\over(D-2)}} \eta_{ab} dx^a dx^b
    + \ell^2 (\frac{\rho_0}{\rho})^{{(D-3)\over(D-2)}}
    e^{(\rho_0^2 - \rho^2)/\rho_0} \Big(d\rho^2 +
d\q^2\Big)~,\label{e:metricrho11}
\end{equation}
is identical to that of the global cosmic brane
solution studied by the authors of Ref.~\cite{cohen}.
  They considered a theory with a complex
scalar coupled to gravity in which the global $U(1)$ is spontaneously
broken. For the particular case of $D=6$ this gives rise to a cosmic
3-brane which naturally can produce an exponential hierarchy between
the electroweak and gravitational scales.
By compactifying $x_4,{\cdots},x_7$ on a small fixed K3 or a
$T^4$ in~(\ref{e:metricrho11}), we have thus  {\it derived\/} this type of
solution from string theory.

Let us now study our solution in more detail following~\cite{cohen}.
First, in hindsight it is not too surprising that our solution
corresponds to a global cosmic brane. Since the Einstein tensor
only depends on $\rho$ while $T_{\mu\nu}$ is independent of $\rho$
they have to be constant in order to
satisfy~(\ref{e:abrho})-(\ref{e:thth}).
This is exactly the starting point of the model considered in~\cite{cohen}.
The location of the brane is at $\rho=\infty$, which corresponds to
$z=0$. (Recall the
change of variables in making the ansatz for the
metric~(\ref{e:metricrho}), $|z|=r_0e^{\rho_0-\rho}$.) 
However, unlike the supersymmetric case, in which
the $D7$-brane is a delta-function source for the stress-energy
tensor, the global cosmic brane has finite extent. Roughly speaking
the size is that of the location of the core, {\it i.e.},
$|z|\sim r_0$ or $\rho\sim\rho_0$. As we will see $\rho_0$ (or
equivalently $r_0$) plays an
important role in relating the string coupling to the size of the
exponential hierarchy.

Eq.~(\ref{e:metricrho11})
has two (naked) curvature singularities, one at the center of the
core, $z=0$ or $\rho=\infty$, and one at $|z|=\ell$ or
$\rho=0$. 
In the latter case the spacetime ends on this naked
singularity which is located at a finite proper distance from the
core of the brane. 
According to
Ref.~\cite{RG}, one can obtain a different solution by adding a small
negative bulk cosmological constant. (This would correspond in our case to
a negative correction to the flat potential on the moduli space.)
At small distances the solution is that of~\cite{cohen} since the
negative contribution to the stress-energy
tensor is negligible. However, the effect
of the cosmological constant is important at large distances
such that the spacetime becomes smooth.
In~\cite{cohen} the authors argue that the singularity at $z=0$ is
unphysical since the solution does not satisfy the properties of a
global cosmic brane; the stress-energy tensor is zero at the center
of the core. To completely solve for the global cosmic brane one
would have to match the above solution~(\ref{e:metricrho11}) to the
metric inside the 
core and to the metric beyond $|z|=\ell$. 
Strictly speaking, we cannot claim to have a global cosmic
brane solution from string theory until this is done. For now,
however, we will not address this issue. We are hopeful that when
string corrections are taken into account the singularities will be
smoothed out~\footnote{One possibility is the enhan{\c c}on
mechanism~\cite{jpp}. In our case, this would
correspond to having zero stress-energy tensor inside the core.}.

We now compute the tension.
For both of our two particular solutions, (\ref{e:tau1})
and~(\ref{e:tau2}),
  the integrand in Eq.~(\ref{e:Etension}) is
${2\over\rho_0^2}\rho$. Thus, upon integration
from the core to the boundary of spacetime, $0\leq \rho\leq \rho_0$,
assuming that the contribution from inside the core is finite, small
and positive, we find 
$E=2\pi + O(e^{-\rho_0})$. From~(\ref{e:metricrho11}), after changing
back to the 
$z$-variable, it is easy to see that the deficit angle, $\delta
\theta=2\pi$. Since $E > \delta \theta$ from Eq.~(\ref{e:BPS})
the solution is not BPS.
Recall the monodromy outside the core given by Eqs.~(\ref{e:tau1tr}) and~(\ref{e:tau2tr}). 
These monodromies would in the
supersymmetric case correspond to $E_n$  and $\hat E_n$ (the affine extension 
of $E_n$)
configurations for $0\leq n\leq 8$~\footnote{From the 
monodromy alone we cannot distinguish between $K_{\tau_{I,II}}$ (the monodromy
matrix for $\tau_{I,II}$) and $-K_{\tau_{I,II}}$. Hence, there are certain 
subtleties as far as identifying the precise $D7$-brane configuration. For now
we will ignore this issue.} 
which consist of $n{+}2$ and $n{+}3$ $D7$-branes respectively~\cite{zwiebach}.
However, these configurations cannot account for the deficit angle 
$\delta \theta = 2\pi$. On the other hand, a combination of
$n{+}2$ (or $n{+}3$) $D7$-branes placed at one, and $10{-}n$ (or $9{-}n$)
$D7$-branes placed at the other 
singularity can produce the deficit angle of $2\pi$. In fact, the
computation of the 
monodromy around the respective singularities tells us that the charges of 
these two collections of $D7$ branes should
be opposite to each other. This indicates the possibility that the
naked singularities could be resolved by replacing them with the above
$D7$-branes, subject to the appropriate boundary
conditions. In particular, the metric~(\ref{e:metricrho11}) would be
matched to the metric of the respective $D7$-brane configurations at
$|z|=r_0$ and $|z|=\ell$.

As noted above,  the solution is by construction non-supersymmetric,
and thus it is not in general protected against string corrections.
However, the particularly interesting feature of this solution is
that the ratio
of the $D{-}2$- and $D$-dimensional Planck scales can be naturally
large, as in Ref.~\cite{cohen}.
Let us therefore briefly review the analysis of \cite{cohen}. Let $v$ be a new
coordinate, $v\equiv v(\rho)$:
\begin{equation}
   v = \ell \int_{0}^{\rho} e^{(\rho_0^2 - {\varrho}^2)/2\rho_0}
    ({\rho_0\over\varrho})^{{(D-1)\over2(D-2)}} d\varrho~
    =~\ell e^{\rho_0/2}\Big({\rho_0\over2}\Big)^{3D-5\over4(D-2)}
     \textstyle\gamma({D{-}3\over4(D{-}2)},{\rho^2\over2\rho_0})~,
      \label{e:vtrans}
\end{equation}
where $\gamma(\a,\xi)\define\int_0^\xi dt\>e^{-t}t^{\a-1}$ is the
incomplete (little) gamma-function. Note that we restrict the
integration to $0<\rho<\rho_0$, {\it i.e.}, where our solution is valid.
Then the metric for our solution
reads
\begin{equation}
    ds^2 = ({\rho\over\rho_0})^{{2\over(D-2)}} \eta_{ab} dx^a dx^b
    +({\rho\over\rho_0})^{{2\over(D-2)}} dv^2
    + \ell^2 ({\rho_0\over\rho})^{{(D-3)\over(D-2)}}
    e^{(\rho_0^2 - \rho^2)/\rho_0}  d\q^2~.\label{e:metricrho12}
\end{equation}
To examine the gravitational effects of the cosmic brane one
considers metric perturbations of the form
\begin{equation}
    ds^2 = ({\rho\over\rho_0})^{{2\over(D-2)}} (\eta_{ab} + h_{ab})
           dx^a dx^b +({\rho\over\rho_0})^{{2\over(D-2)}} dv^2
           + \ell^2 ({\rho_0\over\rho})^{{(D-3)\over(D-2)}}
           e^{(\rho_0^2 - \rho^2)/\rho_0}  d\q^2~.
   \label{e:metricrhol3}
\end{equation}
After imposing the gauge condition $\vd_a h^{ab} = h^a_a =0$ (see
e.g.~\cite{rs2}), we look for 
solutions of the form
\begin{equation}
   h_{ab} = \varepsilon_{ab} e^{i p_a x^a} e^{in \q}
   \frac{\varphi{(v)}}{\psi{(v)}}~, \label{e:gravexp}
\end{equation}
where the polarization tensor $\varepsilon_{ab}=${\it const.},
\begin{equation}
   \psi(v(\rho)) = \ell^{1/2} e^{(\rho_0^2 - \rho^2)/4\rho_0}
    (\frac{\rho}{\rho_0})^{{D-3\over4(D-2)}}~,\label{e:psi}
\end{equation}
and $ \eta_{ab} p^a p^b = - m^2$, as in Ref.~\cite{cohen}.

The linearized Einstein equation for this type of
fluctuations~(\ref{e:gravexp}) reduces to
\begin{equation}
   [- {d^2\over dv^2} + {1\over\psi} {d^2 \psi\over d v^2}
    +n^2  e^{(\rho^2 - \rho_0^2)/\rho_0}
    ({\rho\over\rho_0})^{{D-1\over D-2}}] \varphi (v) = m^2 \varphi (v)~.
\label{e:Sch}
\end{equation}
As shown in Ref.~\cite{cohen}, there is one normalizable zero mode
($n = m =0$), which corresponds to $\f(v) = \psi(v)$, and is well
behaved near the singularity. (There is another zero mode which
diverges near the singularity, which as  argued by the authors of
\cite{cohen}, can be eliminated by choosing appropriate boundary
conditions. These boundary conditions prevent all conserved
quantities from ``leaking out'' through the boundary.)
As pointed out in \cite{cohen} (see also \cite{csaki})
the operator on the left-hand side of Eq.~(\ref{e:Sch})
for $n=0$ is positive semidefinite, and hence $m^2\geq 0$.

Following~\cite{cohen}, the existence of a normalizable zero mode
allows us to obtain the
  relation between the four- and six-dimensional Planck
scales~\footnote{The relation between the six and the ten-dimensional
Planck scales is that of an ordinary Ka{\l}u\.{z}a-Klein
compactification; $M_6^4=M_{10}^8 \mbox{Vol}(M_2)$, where $M_2$ is either K3
or $T^4$.}
\begin{equation}
    M_4^2 = M_6^4 \int \psi^2(v) dv d\q~ =  \pi\ell^2 e^{\rho_0}
    \rho_0^{5/8}\Gamma(\frc{3}{8})M_6^4~. \label{e:scaling}
\end{equation}
If the upper limit of integration is $\rho_0$, $\Gamma(\frc{3}{8})$
should be replaced with $\gamma(\frc{3}{8}, \rho_0)$; however, the
difference quickly becomes negligible if $\rho_0 > 1$.
The exponential dependence of $M_4/M_6$ on $\rho_0$ produces
an exponential hierarchy, except for small $\rho_0$,
where~(\ref{e:scaling}) reverts to a power-law.
Our non-supersymmetric 3-brane solution, derived explicitly from
string theory, hence naturally provides for an exponential
hierarchy. In addition, the exponential hierarchy is naturally related
to a string coupling 
of $O(1)$. To see this, 
recall that in our solutions~(\ref{e:tau1})
and~(\ref{e:tau2}),
  the dilaton is of the form
$\phi=-2\sqrt{2}\theta/\sqrt{\rho_0}$. Unless $\rho_0$ is very small
this means that $\phi\sim\pm O(1)$. In particular, if $\rho_0=8\pi^2$ 
then $M_4/M_6 \sim 10^{18} (\ell M_6)$.

The Fourier $\q$-expansion in Eq.~(\ref{e:gravexp}) is now easy to
recognize as the Ka{\l}u\.{z}a-Klein mode expansion, especially
since it is the $n{=}0$ mode that turns out to correspond to the
four-dimensional graviton. 
The spectrum must be discrete since the volume of the transverse
space is finite, and hence the mass gap is $\delta m^2 \sim
M_6^4/M_2^2$. (For a more detailed analysis, see appendix~A.)
Note that although $\delta m^2\ll1$ (in units of
$M_6^2$) the corrections to Newton's law are of Yukawa potential type
and thus exponentially suppressed.

We now turn to the case of a varying K3 compactified string
theory. (A varying $T^4=T^2{\times}T^2$ is easy to deal with by
iterating the analysis from the previous section.) In this
situation, the non-compact spacetime is $D=6$ dimensional. As
before, the moduli of the K3 depend on
$z{=}x_5{+}ix_6$ and its conjugate, $\bar z$. Rather than studying
the full $20$-dimensional moduli space, we will focus on a
one-parameter family relevant for discussing $A_k$ singularization
of a $K3$ surface. In appendix~B, the metric is worked out as a
function of $t$, the deformation parameter in the defining equation
for an $A_k$ singularity,
\begin{equation}
    XY+U^{k+1}=t(z,\zb)~.
\end{equation}
One can show that, for $|t|<(R')^2$ (see appendix~B), we have
\begin{equation}
   \begin{array}{rcl}
    G_{t\bar t} &=&
    \vd_t\vd_{\bar t} \big(-\log(K(t;2))\\[2mm]
    &=&\Big(2R'^2|t|^{2k\over k+1}K(t;2)\Big)^{-1}+
    \Big(4R'^4|t|^{2(k-1)\over k+1}K^2(t;2)\Big)^{-1}~.
    \label{e:Kt2}
   \end{array}
\end{equation}
This is to be compared with the metric on the moduli space in the
vicinity of a nodal $T^2$, $XY=t(z,\bar z)$, (see appendix~B):
\begin{equation}
   \begin{array}{rcl}
    G_{t\bar t} &=&
    \vd_t\vd_{\bar t} \Big(-\log(-\log(t)-\log(\bar t))\Big)\\[2mm]
    &=&{[t\bar t\log^2 (t\bar t)]}^{-1}~.\\
   \end{array}
\end{equation}

As in the case of the $T^2$, we then proceed to obtain the
equations of motion from the action. We get essentially the
same solution, except that the K\"ahler metric $G_{\a\bar\beta}$
in~(\ref{e:moduli}) is replaced with the one in Eq.~(\ref{e:Kt2}).
Near the location of the singularity, it is enough to keep the leading
term. If we consider the solution for which $\vd_\rho t=0$, where
$\rho=\log r$, $z=re^{i\q}$ as before, and $t=\bar t$, one finds in
the limit $t\to 0$
\begin{equation}
   \vd^2 t - \frac{k}{(k+1)}\frac{1}{t}
   (\vd t)^2 =0~.\label{e:tK3k}
\end{equation}
This can be compared with the case for the torus
\begin{equation}
   \vd^2 t - \frac{1}{t}
   (\vd t)^2 =0~.\label{e:tT2}
\end{equation}
The particular solution of the type~(\ref{e:tau1}) for
Eq.~(\ref{e:tT2}) is of the form $t=\tilde c e^{\tilde w\q}$, while
that of~(\ref{e:tK3k}) is $t=\hat c (\q/(k+1))^{k+1}$. Note that
the latter approaches the former when $k\to \infty$.
Furthermore, when taking $t\to 0$ we are restricting $t$ to be real because
of the way that $t$ is related to $\tau$, as $t\sim\exp(-2\pi
\tau_2)$. The other solution can be dealt with in a similar fashion.
It is worth pointing out that in this case one would expect gauge
  and matter
degrees of freedom on the cosmic brane from wrapped $D$-branes on
the shrinking cycles\footnote{We thank S.~Kachru for reminding us of
this.}.

\section{Discussion}

To summarize, we have shown in this paper that exponential hierarchy can 
naturally arise from non-supersymmetric spacetime varying string vacua.
The emergence of an exponential hierarchy
is naturally
related to having the string coupling of $O(1)$~\footnote{However, we can in 
principle access all of the
coupling space by taking $\rho_0$ very small in which case the dilaton
$\phi$ varies over all of the real line as $\theta$ goes from $-\pi$ to
$\pi$.}.
The four dimensional Planck scale is dynamically determined by the
Planck scale in the bulk and the string coupling.
In the $D=10$ Type~IIB theory, the r\^ole of space-dependent moduli
is played by the dilaton-axion system. This solution, when further
compactified on a $K3$ or $T^4$, leads to a four dimensional 
world previously considered only as a phenomenological solution.
Note that the emergence of exponential hierarchy without demanding
supersymmetry in this particular brane world scenario is akin to
the similar phenomenon in technicolor.

There are many realizations of our scheme besides 
the straightforward compactification on $K3$ or
$T^4$. For example, iterating non-trivially the above analysis, one
might compactify $x_8,x_9$ on a $T^2$ and have its complex structure
modulus, $\tau$, vary over $z{=}x_6{+}ix_7$, while its complexified
K\"ahler class, $\varrho$, varies over $w{=}x_4{+}ix_5$. Each
modulus then gives rise to a 5-brane 
intersecting in a
cosmic 3-brane. Another
possibility is to have three intersecting 7-branes in $D=10$
type IIB theory, repeating the earlier discussion for the individual 7-branes.
Each brane depends on a different transverse complex plane and hence
they intersect in a 3-brane~\footnote{It is interesting to note that
because the branes carry RR-charge open strings stretch from one 7-brane to
another. Since there are three branes this would give rise to three
different types of matter multiplets, which in principle could account
for the three generations observed in nature.}.

In all of our scenarios the issue of stability is clearly very
important. One possibility is that the non-supersymmetric
solutions we have found are
unstable and that they
decay to a supersymmetric set of $(p,q)$ $D7$-branes. It is natural to
expect that
certain properties would be preserved in this decay such as the $SL(2,\ZZ)$
monodromy transformation. DeWolfe et al~\cite{zwiebach} have
computed the monodromy  for all possible types of $(p,q)$ $D7$-brane
configurations. As discussed in section~3, the
$E_n$ ($\hat E_n$) and $H_{8-n}$ ($A_{8-n}$) singularities, given
by sets of $n{+}2$ ($n{+}3$) and $10{-}n$ ($9{-}n$) isolated
$D7$-branes, 
have monodromies, which
coincide with 
the $SL(2,\ZZ)$ transformations
of our $\tau_{I}$ ($\tau_{II}$) solution. 
This could be taken as an indication that our solutions are longlived
excitations
around these supersymmetric vacua. One could imagine that the naked
singularities 
in our metric can be
resolved in the following manner. After imposing appropriate boundary
conditions we 
patch up the spacetime metric of the supersymmetric solutions with
the non-singular part of our metric. The solution obtained in this way
would hopefully have the essential features of our solution, {\it i.e.},
exponential hierarchy.
We hope to return to some of these issues in the
future~\cite{bhmtwo}.

Notice that the warp factors in the expression for  the metric that
describes our solution depend only on one extra dimension. It seems
natural to ask whether there exists a holographic renormalization
group \cite{rg} interpretation of the bulk equations of motion that
describe this brane-world solution.

Finally, it would be interesting to see whether the recent
discussion about the possible relaxation mechanisms for the
cosmological constant~\cite{cc} applies to this particular class of
string vacua. It would be also interesting to
understand whether the conjectured relationship between holography
and the cosmological constant problem \cite{cc1} can be made more
specific in this situation.

{\bf Acknowledgments:}
We thank V.~Balasubramanian, O. Bergman,
R.~Corrado, M.~Dine, E.~Gimon, J.~Gomis,
P.~Ho\v{r}ava, G.~Horowitz,
T.~Imbo, N.~Itzhaki, C.~Johnson, S.~Kachru, P.~Mayr, A.~Peet,
J.~Polchinski, E.~Silverstein,
S.~Sethi, K.~Sfetsos and S.~Thomas for useful discussions.
    The work of P.~B.\ was supported in part by the National Science
Foundation under grant number PHY94-07194. P.B. would like to thank the
Caltech/USC Center for Theoretical Physics,
as well as Stanford University and University of
Durham for their hospitality in the final stages of this project.
    T.~H.\ wishes to thank the US Department of Energy for their
generous support under grant number DE-FG02-94ER-40854 and the
Institute for Theoretical Physics at Santa Barbara, where part of
this work was done with the support from the National Science
Foundation, under the Grant No.~PHY94-07194.
    The work of D.~M.\ was supported in part by the US
Department of Energy under grant number DE-FG03-84ER40168.
D.~M.\ would like to thank Howard University and the University of
Illinois at Chicago for their hospitality while this work was in
progress.

\appendix
\section{Potential for massive gravitational modes} \label{potential}

In this appendix we analyze the potential, ${\psi^{-1}} d^2 \psi/d v^2$,
in the linearized Einstein equation for
the gravitational fluctuations~(\ref{e:Sch}).

Note that
for $\rho\sim0$, we have the power series expansion~\cite{Arfken}
\begin{equation}
       v = \ell\inv2 e^{\rho_0\over2}\rho_0^{~{D-1\over2(D-2)}}\,
       \rho^{D-3\over2(D-2)}\sum_{k=0}^\infty
       {\big({-}\rho^2/2\rho_0\big)^k
        \over k!\Big({D-3\over2(D-2)}+k\Big)}~,\qquad \rho\sim0~,
        \label{e:smallrho}
\end{equation}
so that $\psi \approx (\ell v/\rho_0)^{1/2}$. While an asymptotic
formula for $v=v(\rho)$ may be derived for $\rho\to\infty$, it turns
out to be less reliable as it involves a formally divergent series.

\begin{figure}[ht]
    \epsfxsize=110mm%
    \hfill\epsfbox{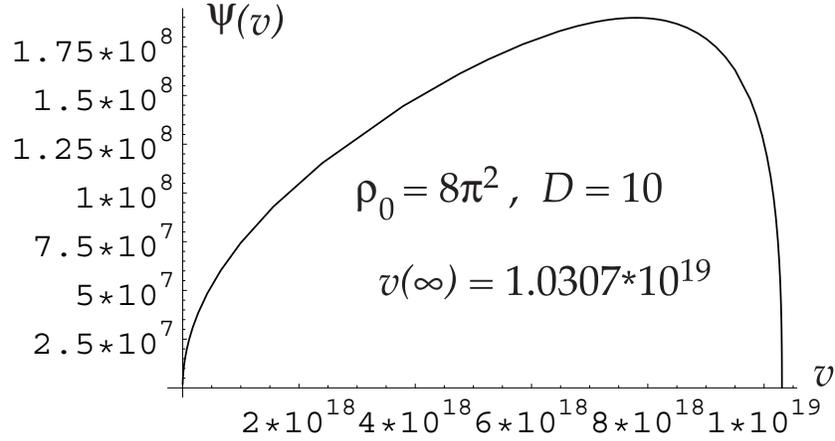}\hfill~\\
    \caption{A parametric plot of $\psi(\rho)$, as given in
     Eq.~(\ref{e:vtrans}), {\it vs.} $v(\rho)$, as given in
     Eq.~(\ref{e:psi}). Note that $v$ and $v_\infty$ are
given in units of $\ell$.}
    \label{f:psiV0}
   \end{figure}

Since the transcendental change of variables~(\ref{e:vtrans}) is not
invertible in closed form, we are unable to give a $\psi$ as an
explicit function of $v$ in closed form. However, using
{\sl Mathematica}'s {\tt ParametricPlot}, we can plot $\psi$ {\it
vs.} $v$, and the result is shown in Fig.~\ref{f:psiV0}.
Clearly, as a function of $v$, $\psi(v)$ is well approximated by
\begin{equation}
    \psi(v;c)\define\sqrt{v}(\frac{v_\infty{-}v}{\ell})^{1/c}~,\label{e:psi8}
\end{equation}
for a suitable $c$.

\begin{figure}[ht]
    \epsfxsize=120mm%
    \hfill\epsfbox{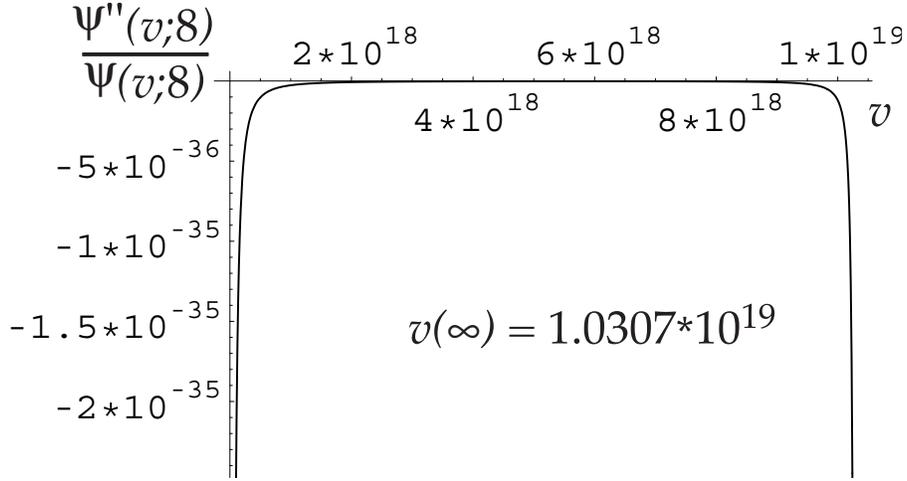}\hfill~\\
    \caption{A plot of the potential approximating $\psi''/\psi$,
             as given in Eq.~(\ref{e:pot}), and using $c{=}8$ in
             Eq.~(\ref{e:psi8}). Note that $v$ and $v_\infty$ are
given in units of $\ell$.}
    \label{f:pot}
   \end{figure}

With this approximation, we obtain
\begin{equation}
   {1\over\psi}{d^2\psi\over dv^2} \approx
    -{1/4\over v^2} - {1/c\over v(v_\infty{-}v)}
    - {(c{-}1)/c^2\over(v_\infty{-}v)^2}~,
    \label{e:pot}
\end{equation}
which is plotted in Fig.~\ref{f:pot}.

Note that these terms may be thought of as the superposition of
two attractive potentials, one at $v{=}0$ and the other at
$v{=}v_\infty$. The attractive potential at $v{=}0$ (and so
$\rho{=}0$, {\it i.e.}, $|z|=\ell$) corresponds to a naked singularity a
finite distance from the core at $\rho=\rho_0$ (or $|z|=r_0$),
while the one
at $v_\infty$ ({\it i.e.}, $\rho=\infty$) corresponds to the other
singularity at the origin of the $z$-plane.

The approximate potential may well define an exactly soluble quantum
mechanical problem. For now, however, let us discuss a further
simplification. Firstly, we neglect the middle term in
Eq.~(\ref{e:pot}), since it is dominated by the first and the last
term, when $v\sim0$ and $v\sim v_\infty$, respectively. Secondly,
except for the `charge' of the potential terms, $-\inv4$ for the
first term and $-{c{-}1\over c^2}$ for the last, the two terms
represent the same type of divergence. Thus we analyze `the first
half of the potential', {\it i.e.}, $V(v)=-\inv{4\,v^2}$, and look
for solutions of the Schr\"{o}dinger equation
\begin{equation}
   \Big[-{d^2\over dv^2}-{1\over4\,v^2}\Big]\varphi^{(0)}
    =m^2\varphi^{(0)}~,
   \label{e:simple}
\end{equation}
of the form $\varphi^{(0)}(v)=\sqrt{v}f(v)e^{-bv}$. Upon
substitution, we find
\begin{equation}
    \Big[{1\over4\,v^2}-{f'\over vf}+{b\over v}-{f''\over f}
        +{2bf'\over f}-b^2-{1\over4\,v^2}-m^2\Big]\varphi^{(0)}=0~.
\end{equation}
This suggests the identification $b=\sqrt{-m^2}$: for negative
`energy', $m^2{<}0$, $b$ is real and the solutions would
exponentially decay; for positive `energy', $m^2{>}0$, $b$ is
imaginary and the solutions would have a plane wave factor. The
resulting equation for $f(v)$ then is
\begin{equation}
   vf'' + (1-2bv)f' - bf = 0~,
\end{equation}
which is easily solved in terms of a power series,
$f=\sum_{k=0}^\infty c_k\,v^{k+s}$. Direct substitution yields
\begin{eqnarray}
   0&=&\sum_{k=0}^\infty c_k(k+s)[(k+s-1)+1]v^{k+s-1}
      -b\sum_{k=0}^\infty c_k[2(k+s)+1]v^{k+s-1}~, \nn\\
    &=&c_0(s)^2v^{s-1}
    +\sum_{k=0}^\infty\Big\{c_{k+1}(k+s+1)^2
      -c_kb[2(k+s)+1]\Big\}v^{k+s}~.
\end{eqnarray}
The vanishing of the coefficient of the first term implies $s{=}0$,
and the vanishing of the coefficients for the remaining terms imply
the recursion relation:
\begin{equation}
    c_{k+1} = c_k\,b\>{2(k+s)+1 \over (k+s+1)^2}~. \label{e:rec}
\end{equation}
Since
\begin{equation}
   {c_{k+1}\,v^{k+1} \over c_k\,v^k}={2k+1 \over (k+1)^2}\,bv
   \to{2bv\over k}<1~,\quad\mbox{when}\quad k\to\infty~, \label{e:lim}
\end{equation}
the series defined by the recursion relation~(\ref{e:rec}) converges.
Moreover, since large powers of $v$ dominate when $v$ is large, we
have that $f(v)\sim e^{2bv}$ when $v$ is large, since the power
series of $e^{2bv}$ has the limiting ratio~(\ref{e:lim}). Note that
the asymptotically exponential dependence ensures that $v$ equaling
a small fraction of $v_\infty$ is already sufficiently large for the
asymptotic estimate (see Fig.~\ref{f:pot}). Thus, although the
present solution is derived for the simplified toy potential in
Eq.~(\ref{e:simple}), it will turn out to be useful both for
the full toy model, with the potential~(\ref{e:pot}), and also for
the real case, with the potential $\psi''/\psi$.

As there is no choice of $b=\sqrt{-m^2}$ for which the series
would terminate, we conclude that
$\varphi^{(0)}(v)=\sqrt{v}f(v)e^{-bv}\sim\sqrt{v}e^{+bv}$ for large
$v$. For $m^2{<}0$, when $b$ is real and positive, all such
solutions are unphysical as they are unnormalizable. (Recall that on
retaining only the simple potential as shown in~(\ref{e:simple}), we
also let $0\leq v\leq\infty$.) For $m^2{>}0$, however, $b$ is
imaginary, and
$\varphi^{(0)}(v)=\sqrt{v}f(v)e^{-i\beta v}\sim\sqrt{v}e^{i\beta v}$,
where $\beta=|b|$. Note in particular that $\varphi^{(0)}(0)\equiv0$.

Since the potential~(\ref{e:pot}) is qualitatively symmetric with
respect to $\{v{=}0\}\leftrightarrow\{v{=}v_\infty\}$, we expect to
find appropriately `reflected' solutions around the attractive
potential at $v{=}v_\infty$. Based on the result in the previous
case, we also expect
$\varphi^{(\infty)}(v)=(v_\infty{-}v)^{1/c'}g(v)e^{-b(v_\infty-v)}$,
and in particular that $\varphi^{(\infty)}(v_\infty)=0$.

Then, the solutions for the full toy potential~(\ref{e:pot}), and also
for our actual problem~(\ref{e:Sch}) with $n{=}0$, should be well
approximated by the linear superpositions
$\varphi^{\pm}\define\inv{\sqrt2}
   (\varphi^{(0)}\pm\varphi^{(\infty)})$. Easily,
$\varphi^{\pm}(0)\equiv0\equiv\varphi^{\pm}(v_\infty)$---as if there
existed impenetrable walls at $v{=}0$ and $v{=}v_\infty$. That is,
the quantum mechanical problem with the toy potential~(\ref{e:pot}),
and also our actual problem~(\ref{e:Sch}) with $n{=}0$, behave
{\it qualitatively\/} as if the wave-functions $\varphi(v)$ were
confined in a (smoothed) infinite potential well. Therefore, we
expect a discrete `energy' spectrum $m^2{\geq}0$, and a mass
gap between $\psi$, the $n{=}0{=}m$ solution of Eq.~(\ref{e:Sch}) and
the lowest lying state with $m^2{>}0$.

\section{K\"ahler Potential Singularities} \label{WPZ}
In this appendix we study the singularities of the
Weil-Petersson-Zamolod\-chi\-kov K\"ahler potential on the complex
structure moduli spaces for several cases of interest. This follows
and extends the results of Ref.~\cite{Roll}, using the hallmark of
Calabi-Yau $n$-folds, $M_n$: their covariantly constant and nowhere
vanishing $(n,0)$-form. Variations of this $(n,0)$-form describe the
complex structure moduli space, in which a codimension-1 subspace
(the discriminant locus) corresponds to singular $n$-folds.

\subsection{The potential near singularities}
Following Ref.~\cite{Philip}, the holomorphic $(n,0)$-form may be
written as
\begin{equation}
\W_t = {\prod_{i\neq j}dx_i \over \f,_j(x;t)}~, \qquad
    i,j=1,{\cdots},n{+}1~.
\end{equation}
Here, $x_i$ are local coordinates on the $n{+}1$-fold, $\cal Y$, in
which the Calabi-Yau $n$-fold, $M_n$, is embedded as the
hypersurface $f(x;t)=0$; the $t$ are local coordinates on the
(complex structure) moduli space. Generalizations of this to
complete intersections of hypersurfaces and other elements of
cohomology are straightforward~\cite{Philip,Res}.

The integral $K(t;n)\define i^n\int_{M_n}\ba{\W}_t{\wedge}\W_t$ turns
out to play a double r\^ole in the models we consider. Its logarithm
is the K\"ahler potential for the Weil-Petersson-Zamolodchikov metric
on the complex structure moduli space~\cite{Roll,WPZ}. It is also the
relative conformal factor in the spacetime metric for the line element
transversal to the cosmic $(7{-}2n)$-branes~\cite{gh}; the peculiar
factor, $i^n$, ensures the Hermiticity of $K(t;n)$, as necessary for
its latter r\^ole.

As a toy model, compactify the spacetime coordinates $x_8,x_9$ on a
$T^2$, the complex structure of which is permitted to vary over the
space-like $(x_6,x_7)$-surface. To this end, $M_1=T^2$ may be
defined as a cubic hypersurface $f(y;t)=0$ in ${\cal Y}=\CP2$, the
coefficients of which are functions of the moduli, $t$, which in
turn are functions of $z{=}x_8{+}ix_9$ and its conjugate:
\begin{equation}
\begin{array}{rcl}
f(y;t)&\define&
y_1^{~3}+y_2^{~3}+y_3^{~3}-3\,t(z,\zb)\,y_1y_2y_3~=~0~,\\[2mm]
&&(y_1,y_2,y_3)\simeq(\l y_1,\l y_2,\l y_3)~.\\ \end{array}
\label{e:tT}
\end{equation}
This torus becomes singular ($d_yf=0$) at $t_n=e^{2n\p i/3}$,
$n=0,1,2$, and $t_\infty=\infty$. At each of these points in the
$t$-plane, and the corresponding points of the $z$-plane, this
highly symmetric torus has three singular points, each of which
may be described as a node, {\it i.e.}, an $A_1$ singularity in
the classification of Ref.~\cite{Arnold}.
That is, first change variables in the $t$-plane so that a
singularity of Eq.~(\ref{e:tT}) is moved to $t=0$. Then, a
holomorphic (but nonlinear) change of local coordinates on $\CP2$
turns the defining equation, $f(y;t){=}0$, into $XY=t$, up to
higher order terms\footnote{For example, let $t\to1/3t$, work in the
coordinate patch where $y_3\neq0$, divide through by $y_3^{~3}$,
define $X=y_1/y_3$ and $Y=y_2/y_3$ and neglect $X^3,Y^3$ as compared
to 1.}. For nonzero $t$, the solution of this equation is a
rotational hyperboloid. As $t\to0$, this pinches into a two-sheeted
cone with the vertex---the singularity---at $X{=}0{=}Y$. The
holomorphic $(1,0)$-form may here be written~\cite{Philip,Roll} as
\begin{equation}
   \W_t = f(Y){dY\over Y}~,\qquad \bar{\vd}_Y\,f(Y)=0~,
   \label{e:Omega1}
\end{equation}
this expression being valid within the neighborhood $|X|,|Y|<R$.
Furthermore, since $X=t/Y$, we have
\begin{equation}
   R>|X|={|t|\over|Y|}~,\qquad\To\qquad |Y|>{|t|\over R}~,
\end{equation}
so that ${|t|\over R}<|Y|<R$, and therefore also $|t|<R^2$.

The integral $i\int_{M_1}\ba{\W}_t{\wedge}\W_t$, as a function of
$t$, may then be estimated by dividing the integral into the
contributions from each neighborhood in which a singularity
develops, and the remaining part which is regular in $t$. The
singularities all being equal, locally, we calculate the
contribution from the one at $t{=}0$ and multiply by the number of
them, $N$. Therefore:
\begin{equation}
   K(t;1)\define i\int_{M_1}\ba{\W}_t{\wedge}\W_t = V_0(t)+
   iN\int_{|t|/R<|Y|<R}\big|f(Y)\big|^2{dY{\wedge}d\bar{Y}\over|Y|^2}~.
   \label{e:IntBeg1}
\end{equation}
Following Ref.~\cite{Roll}, we expand $f(Y)$ in a Taylor series, and
note that all the positive powers integrate to regular functions of
$t$. We may therefore write
\begin{equation}
   i\int_{M_1}\ba{\W}_t{\wedge}\W_t = V(t) +
   2N|f_0|^2\int_{|t|/R<|Y|<R}{d^2Y\over|Y|^2}~,
\end{equation}
where all the higher order contributions from $[f(Y){-}f_0]$ are
absorbed in the regular function $V(t)$. Writing $Y=re^{i\q}$, we
easily obtain:
\begin{equation}
\int_{|t|/R<|Y|<R}{d^2Y\over|Y|^2}
    = \int_0^{2\p}d\q\int_{|t|/R}^R{rdr\over r^2}
    = 2\p\log\big({R^2\over|t|}\big) \label{e:IntFin1}
\end{equation}
Upon the transformation $t=R^2j(\t)\sim R^2e^{2\p i\t}$ near $t=0$,
{\it i.e.}, near $\Imm(\t)=\infty$,
\begin{equation}
   K(t;1)\define
   i\int_{M_1=T^2}\ba{\W}_t{\wedge}\W_t \propto \Imm(\t)~,
\end{equation}
which is indeed the standard K\"ahler potential in the
Teichm\"{u}ller theory~\footnote{Recall that for the Calabi-Yau 1-fold, 
the 2-torus $M_1=T^2$, the
space of complex structures may be parameterized as the $|\t|\geq1$
portion of the $-\inv2\leq\Ree(\tau)\leq+\inv2$ strip in the upper
half-plane $\Imm(\t)>0$.}.

Before we turn to higher dimensional cases, it is useful to consider
the action in terms of the local coordinate $t\propto j(\t)$. This
will facilitate comparison, as
we will momentarily see, with the metric for the $A_k$ singularities.
The K\"ahler metric in the $t$ coordinate near $t=0$ is given by
\begin{equation}
   \begin{array}{rcl}
   G_{t\bar t} &=&
   \vd_t\vd_{\bar t}\Big(-\log[-\log(t)-\log(\bar t)]\Big)\\[2mm]
   &=&\Big(|t|\log|t|^2\Big)^{-2}~.\\
   \end{array}
\end{equation}
Recall that the metric in terms of the $\tau$ coordinate is given by
$G_{\tau \bar{\tau}} = - [\Imm(\tau)]^{-2}$.

We now repeat the calculation of
$K(t;2)\define \int_{K3}\W_t{\wedge}\ba{\W}_t$, for the
Calabi-Yau (complex) 2-folds, the K3 surfaces. The first marked
distinction is that there can now be many more types of
singularities~\cite{Arnold}. Herein, we consider the $A_k$ type, for
which the defining equation may be brought into the general
form\footnote{In Refs.~\cite{vafa,gh}, the modulus $t(z,\zb)$ has
been chosen to be a holomorphic---moreover linear---function of $z$.
This need not be so in general, and indeed our main solutions are
non-holomorphic.}
\begin{equation}
    XY+U^{k+1}=t(z,\zb)~,\label{e:Ak2}
\end{equation}
within the neighborhood $|X|,|Y|,|U|^{k+1\over2}<R$.
Solving for $X$, we have that
\begin{equation}
    R>|X|={|t-U^{k+1}|\over|Y|}~,\qquad\To\qquad
    |Y|>{|t-U^{k+1}|\over R}~,\label{e:lBound}
\end{equation}
which gives a lower bound for $|Y|$. The upper bound, $|Y|<R$, then
implies through Eq.~(\ref{e:lBound})
\begin{equation}
    {|t-U^{k+1}|\over R}<|Y|<R~,\qquad\To\qquad |t-U^{k+1}|<R^2~,
\end{equation}
which is not guaranteed by $|U|^{k+1\over2}<R$ when $t{\neq}0$. We
thus take $|U|^{k+1\over2}<R'$, where $R'$ is sufficiently smaller
than $R$. Therefore, with $N$ isolated singular points of the $A_k$
type,
\begin{equation}
    K(t;2) = V(t) + 2N|f_0|^2\int_{|U|^{k+1\over2}<R'}d^2U
    \int_{|t-U^{k+1}|/R<|Y|<R}{d^2Y\over|Y|^2}~.
\end{equation}
The $Y$-integral straightforwardly gives
$2\p\log\big({R^2\over|U^{k+1}-t|}\big)$, and we are left with the
$U$-integral:
\begin{equation}
    K(t;2) = V(t) + 4\p N|f_0|^2\int_{|U|^{k+1\over2}<R'}d^2U~
    \log\Big({R^2\over|U^{k+1}-t|}\Big)~.
\end{equation}
The result of the $U$-integration depends on whether $|U|^{k+1}<|t|$
or $|U|^{k+1}>|t|$. In the former case, we write
\begin{equation}
    |U|^{k+1}<|t|~:\qquad
    \log\big(|U^{k+1}-t|\big) =
    \log|t| + \log\Big|1-{U^{k+1}\over t}\Big|~,\label{e:tBig}
\end{equation}
while in the latter case we write
\begin{equation}
    |U|^{k+1}>|t|~:\qquad
    \log\big(|U^{k+1}-t|\big) =
    (k{+}1)\log|U| + \log\Big|1-{t\over U^{k+1}}\Big|~.\label{e:uBig}
\end{equation}
In both cases, the second logarithm leads to an integral of the
general form ($w=\rho e^{i\q}$)
\begin{eqnarray}
    &&\hspace{-15mm}\int_{|w|<C} d^2w\log\big|1-aw^b\big|~,
    \qquad |aw^b|<1~, \nn\\
    &=& \inv2\int_0^{2\p} d\q\int_0^C\rho d\rho~
    \Big[\log(1-a\rho^be^{ib\q})
        +\log(1-\bar{a}\rho^be^{-ib\q})\Big]~, \nn\\
    &=& \inv2\int_0^C\rho d\rho\int_0^{2\p} d\q~
    \Big[\sum_{n=1}^\infty{1\over n}a^n\rho^{nb}e^{inb\q}
    +\sum_{n=1}^\infty{1\over n}\bar{a}^n\rho^{nb}e^{-inb\q}\Big]
    ~=~0~.\label{e:Zero}
\end{eqnarray}

Now, if $R'<|t|^{1\over2}$, then also
$|U|^{k+1\over2}<R'<|t|^{1\over2}$, and we obtain:
\begin{eqnarray}
    &&\hspace{-10mm}
     \int_{|U|^{k+1\over2}<R'}d^2U~
      \log\Big({R^2\over|U^{k+1}-t|}\Big)\nn\\
    &=&\int_{|U|^{k+1\over2}<R'}d^2U~\bigg(
      \log(R^2)-\log|t|-\log\Big|1-{U^{k+1}\over t}\Big|\bigg)~,\nn\\
    &=&\p(R')^{4\over k+1}\log\Big({R^2\over|t|}\Big)~,\qquad
      (R')^2<|t|~,
\end{eqnarray}
where we used Eqs.~(\ref{e:tBig}) and~(\ref{e:Zero}) for the second
and the third equality.

On the other hand, if $|t|<(R')^2$, then for part of the
integral $|U|^{k+1\over2}<|t|^{1\over2}<R'$, and in the remaining
part $|t|^{1\over2}\leq|U|^{k+1\over2}<R'$. Splitting the integration
accordingly, we obtain:
\begin{eqnarray}
&&\hspace{-10mm}
    \int_{|U|^{k+1\over2}<R'}d^2U~
     \log\Big({R^2\over|U^{k+1}-t|}\Big)\nn\\
    &=&\int_{|U|^{k+1\over2}<R'}d^2U~\log(R^2)
    -\int_{|U|^{k+1\over2}<R'}d^2U~\log\big|U^{k+1}-t\big|~, \nn\\
    &=&\p(R')^{4\over k+1}\log(R^2)
      -\int_{|U|^{k+1\over2}<|t|^{1\over2}<R'}d^2U~\bigg(
        \log|t|+\log\Big|1-{U^{k+1}\over t}\Big|\bigg)\nn\\
    &&-\int_{|t|^{1\over2}\leq|U|^{k+1\over2}<R'}d^2U~\bigg(
       \log|U|^{k+1}+\log\Big|1-{t\over U^{k+1}}\Big|\bigg)~, \nn\\
    &=&2\p(R')^{4\over k+1}\log(R)
        -\p|t|^{2\over k+1}\log|t|-(k{+}1)2\p\Big[
                   \inv2|U|^2\log|U|-\inv4|U|^2
                    \Big]_{|t|^{1\over k+1}}^{(R')^{2\over k+1}}~,\nn\\
    &=&\frc{k{+}1}2\p\Big((R')^{4\over k+1}-|t|^{2\over k+1}\Big)
    +2\p(R')^{4\over k+1}\log\Big({R\over R'}\Big)~, \quad |t|<(R')^2~,
\end{eqnarray}
where we used Eqs.~(\ref{e:tBig}) and~(\ref{e:uBig}) in the second,
and~(\ref{e:Zero}) in the third equality.

To summarize,
\begin{equation}
    K(t;2) = V(t) + 4\p^2N|f_0|^2 (R')^{4\over k+1} I(t;R,R')~,
    \label{e:K(t;2)}
\end{equation}
where $t=t(z,\zb)$ and
\begin{equation}
    I(t;R,R') = \left\{\!\!\!
    \begin{array}{ll}
    \log\Big({R^2\over|t|}\Big)~,
    &|t|>(R')^2~,\\[3mm]
    {k{+}1\over2}\Big[1-\Big({|t|\over(R')^2}\Big)^{2\over k+1}\Big]
    +2\log\Big({R\over R'}\Big)~,~  &|t|<(R')^2~,\\
    \end{array}\right.\label{e:I(t;R,R')}
\end{equation}
Notice that this function is continuous at $|t|=(R')^{k+1}$, albeit
not smooth.

\end{document}